\documentclass[linenumbers,12pt]{spieman}  
\usepackage[numbers]{natbib}
\usepackage{amsmath,amsfonts,amssymb}
\usepackage{graphicx}
\usepackage{setspace}
\usepackage{tocloft}
\usepackage{natbib}
\usepackage{comment}
\usepackage{lineno}
\usepackage{aas_macros}

\makeatletter
\def\NAT@force@numbers{}
\makeatother

\def\rev#1{\color{black} #1 \rm \color{black}}

\title{Detecting water ice and vapor disks originating from icy planetary bodies around white dwarfs with future PRIMA observations}

\author[a,*]{Ayaka Okuya}
\author[a]{Hideko Nomura}
\affil[a]{National Astronomical Observatory of Japan, 2-21-1 Osawa, Mitaka, Tokyo 181-8588, Japan}

\cftpagenumbersoff{figure}
\cftpagenumbersoff{table} 
\begin{document} 
\maketitle

\begin{abstract}
Observations of atmospheres of polluted white dwarfs provide insights into the elemental composition of accreted exoplanets and exo-asteroids.
However, they poorly constrain the abundance of ice-forming volatile elements due to the properties of white dwarf atmospheres. 
Instead of focusing solely on atmospheric observations, we propose observing circumstellar water ice and vapor disks formed by the tidal disruption of icy bodies using the future PRobe far-Infrared Mission for Astrophysics (PRIMA) far-infrared enhanced survey spectrometer.
PRIMA has the potential to measure volatile abundances in colder circumstellar regions inaccessible by shorter-wavelength observations.
We employ a simple disk emission model with disk parameter ranges inferred from previous observations and disk evolution simulations.
We find the 44-$\mu$m water ice feature promising for observing icy disks. For white dwarfs within 60 pc, 1-hour PRIMA observations could detect water ice with a mass above $10^{20}$ g, representing a potential lower limit of circumstellar disk mass. Water vapor rotational lines also abundantly emerge within the PRIMA wavelength coverage, and 5-hour observations for white dwarfs within 20 pc could detect water vapor with a total disk mass $\gtrsim 10^{20}$ g, depending on the H$_2$/H$_2$O ratio.
19 metal polluted white dwarfs within 20 pc and 210 within 60 pc could be optimal targets for water vapor and ice observations, respectively.

\end{abstract}


{\noindent \footnotesize\textbf{*}Ayaka Okuya,  \linkable{ayaka.okuya.astro@gmail.com} }

\begin{spacing}{1}   

\section{Introduction}
\label{sect:intro}  
Constraining the bulk compositions of planets is crucial for understanding their formation and evolution history. The bulk composition also plays a key role in shaping planetary surface environments and determining their potential habitability. 
The most common approach to inferring the solid bulk composition of an exoplanet is to use its mass--radius relationship \citep[e.g.,][]{Fortney+2007,Zeng+2013,Zeng+2019,Lissauer+2023}. However, this single quantity alone cannot uniquely constrain the composition \citep[e.g.,][]{Seager+2007,Rogers&Seager10,Dorn+15}. To overcome this limitation, the transmission spectroscopy of dusty tails produced by evaporated materials from disintegrating exoplanets has been proposed \cite{Bodman+2018,Okuya+2020}. This method allows for direct probing of the elemental compositions of these planets, but there are currently only three known disintegrating planets \citep{Rappaport+2012,Rappaport+2014,Sanchis-Ojeda+2015}.

White dwarfs (WDs) showing signs of accreted planetary material could allow direct measurements of exoplanetary compositions, with over 1600 such systems \rev{within $\lesssim 1000$ pc} identified to date. Gravitational settling of metals (elements heavier than He) into the deep interior makes WD atmospheres metal-free within a short timescale \citep{Paquette1986}. Nevertheless, metals are detected in 25-50\% of WDs \citep{Zuckerman+2003,Zuckerman+2010,Koester+2014}, suggesting accretion from planets and/or asteroids that survived the post-main-sequence evolution of their host stars \citep{Debes+2002,Bonsor+2011,Jura&Young2014}. Therefore, the elemental abundances in these polluted WD atmospheres likely reflect those of the most recently accreted planetary bodies.

Constraining the abundances of volatile elements, in addition to refractory elements, would enable investigations into planet formation across a broader temperature range, corresponding to a wider orbital region in the protoplanetary disk  \citep{Marboeuf+2014,Eistrup+2018,Harrison+2018}.
The abundances of rock-forming refractory elements such as Mg, Si, Fe, Ca, and O in WD atmospheres have been measured in detail. However, those of ice-forming volatile elements such as H, C, N, (O,) and S remain poorly constrained due to several challenges. 
For instance, the abundance of hydrogen can only be measured in DB-type WDs with a primarily He atmosphere, but it is indistinguishable from primordial hydrogen originating in the stellar atmosphere \citep{Koester+2015,Gentile-Fusillo+2017}. Additionally, detecting oxygen from water requires simultaneous measurement of refractory elements to distinguish it from oxygen in oxides \citep{Farihi+2013}. 
The detection of C, N, and S is biased towards WDs with high-temperature atmospheres whereas Mg, Fe, and Ca are detected across a wide range of atmospheric temperatures \citep{Williams+2024}.

To achieve the ultimate goal of measuring the full elemental composition of accreted planetary bodies, we propose probing volatile elements through observations of circumstellar debris disks around WDs.
According to the planetary accretion scenario, planetary bodies are scattered toward WDs by the gravitational perturbations of nearby large/giant planets and are subsequently tidally disrupted \citep{Debes+2002, Bonsor+2011,Mustill+2018}. 
Fragments produced by the tidal disruption could potentially form compact accretion disks \citep{Jura2003, Veras+2014, Veras+2015, Li+2021, Malamud+2020a,Malamud+2020b,Brouwers+2022}. Infrared emission from hot silicate dust and emission lines of vapor of refractory elements, such as Mg, Fe, and Ca, have indeed been detected around metal-polluted WDs \citep[e.g.,][]{Zuckerman+1987,Jura+2007,Gansicke+2006, Rocchetto+2015,Manser+2016}. If ice-bearing bodies are tidally disrupted, icy particles and volatile gases (e.g., water vapor) could be supplied.
Such volatile gas may experience suppression of accretion due to angular momentum exchange with dust \citep{Okuya+2023}. Additionally, gas that diffuses outward beyond the disk's ice sublimation line might recondense into ice particles, serving as a potential mass reservoir of ice.

Water ice and vapor produce a number of features across near- to far-infrared wavelengths \citep[e.g.,][]{Warren+2008,Rothman+2005,Schoier+2005,Gordon+2022}.
\rev{In particular, the far-infrared ice features appear in emission and trace the total ice column density, whereas near-infrared features, such as the 3 $\mu$m water band accessible with JWST, can only be detected in absorption against a background continuum.}
G29-38 is the only system observed at far-infrared wavelength up to $38~\mu$m so far, suggesting that, in addition to hot rocky dust, cold dust may also be present in order to explain its emission spectra \citep{Reach+2009}. Interestingly, this cold dust component exhibits the continuum slope potentially leading up to the peak of water ice absorption features \rev{(see Fig.~\ref{fig:SED_PRIMA_Ice})}.
On the other hand, among the seven systems with spectra observed below 14 $\mu$m, including G29-38, no gas features from volatile materials have been detected \citep{Reach+2009, Jura+2007AJ, Jura+2009AJ}. Thus, in addition to the observations focused on warmer regions so far, far-infrared spectroscopy is crucial for probing gas and dust in colder regions. The PRobe for-Infrared Mission for Astrophysics (PRIMA), a candidate space mission planned for launch in the 2030s \citep{Moullet+2023}
would enable such observations. Its proposed instrument, the far-infrared enhanced survey spectrometer (FIRESS) \rev{\citep{Bradford+2025}}, will allow low- to high-resolution spectroscopy from 24 to 235 $\mu$m, the wavelengths that have so far remained unobserved in WD disks.

\rev{
Other solid-state features are also abundant in the PRIMA FIRESS band.
For instance, crystalline pyroxenes exhibit strong peaks at 28, 37, and 43 $\mu$m, while crystalline olivines show prominent features at 23, 33, and 69 $\mu$m \citep[e.g.,][]{Fabian+2001, Jaeger+1998}.
In particular, the peak positions of crystalline olivines are known to shift linearly with the Mg/Fe ratio \citep[e.g.,][]{Koike+2003}.
In contrast, amorphous silicates, which are interpreted as the dominant contributors to the 10 $\mu$m silicate feature in previously observed spectra, exhibit an additional spectral feature only at 18 $\mu$m \citep{Jura+2009AJ,Reach+2009}.
Therefore, PRIMA will be able to constrain the fraction of crystalline silicates and determine their detailed compositions.
Additionally, carbonates, tentatively detected in recent JWST spectroscopy \citep{Swan+2024}, exhibit features in the $30-100\ \mu$m range \citep{Posch+2007}.
}

\rev{Furthermore, studying debris disks around WDs can complement studies of those around main sequence (MS) stars, particularly in tracing the chemical evolution of planetary material over cosmic time.
Debris disks around MS stars are more frequently detected around massive and young stars ($<$ Gyr) \citep{Mizuki+2024}, whereas those around WDs are observed in systems with ages of several Gyr, long after their main sequence phase \citep{Koester+2014}. 
Consequently, MS debris disks can be used to reveal the compositions of planetary material that formed relatively recently, within the past $\sim1$ Gyr. In contrast, WD debris disks provide insights into planetary compositions formed in a much earlier epoch, as the material originates from planetary bodies that are thought to have formed before or during their host stars' MS phase.  
In addition, a key advantage of studying WD debris disks is the ability to directly compare the mineralogy of disk dust with the elemental abundances of accreting planetary material measured from WD atmospheres (e.g., Okuya et al. submitted to ApJ) 
}

In this paper, we investigate the feasibility of detecting water ice features and associated volatile gas emission spectra from the disks around WDs by the future PRIMA mission. Using a simple disk emission model, we produce synthetic spectra for both water ice and volatile gas. For the ice disk, we quantify detectability as a function of disk mass and distance from Earth. For the gas disk, we independently vary numerous disk parameters to predict spectra, with plans to develop a photochemical model in a future study to handle each parameter consistently and enable analysis with a minimal parameter set. We demonstrate that PRIMA will be useful for the first detection of water ice and associated volatile gas in the WD disks  across a plausible parameter range.

The organization of this paper is as follows. In Sec.~\ref{sec:method}, we provide the general picture of the disk based on theoretical insights and describe our emission model for dust, ice, and gas disks. In Sec.~\ref{sec:results}, we present the synthetic infrared emission spectra of ice and gas disks and compare them to the expected observational noise of PRIMA/FIRESS. In Sec.~\ref{sec:discussion}, we discuss uncertainties in our dust disk model and candidate observation targets. Finally, we summarize our findings in Sec.~\ref{sec:conclusion}.

\section{Materials and Methods} \label{sec:method}
Using a simplified disk model with plausible parameter ranges, we produce synthetic observation spectra. In Sec.~\ref{subsec:disk-overview}, we outline the general picture of the disk based on theoretical insights. We describe our emission model and parameter settings for ice and gas in Sec.~\ref{subsec:dust-emission} and in Sec.~\ref{subsec:gas-emission}, respectively. In Sec.~\ref{subsec:noise}, we explain the estimation of observational noise for both ice and gas disk observations.

\subsection{Overview of Water Ice and Volatile Gas Disks \label{subsec:disk-overview}}
The icy debris would be supplied by the tidal disruption of icy bodies (mixtures of ice and rock) within the Roche limit ($r_{\rm Roche} \sim 1-3 R_\odot$).
Around the orbit at $r_{\rm Roche}$, the equilibrium temperature is $\sim 600$ K, where rock exists as dust while ice should sublimate into water vapor. In fact, the observed infrared spectra are consistent with the hot silicate dust within the Roche limit and providing constraints on its disk location, mass, and particle size. However, water vapor and ice disks properties remain observationally unconstrained.

\begin{figure}[htbp]
  \centering
  \includegraphics[width=\textwidth]{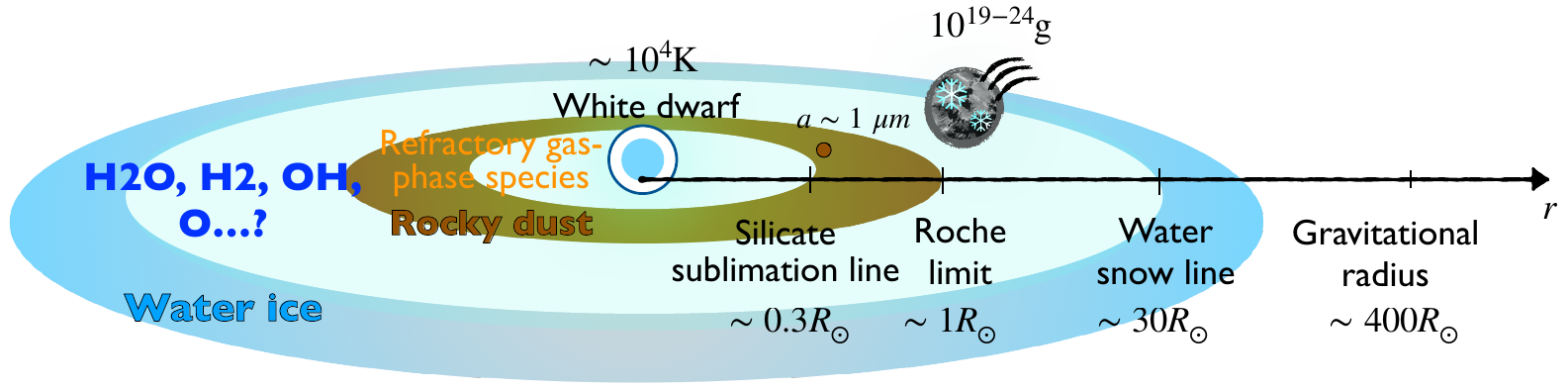}
  \caption{
    Schematic showing the distribution of the modeled volatile gas (e.g., water vapor) disk and water ice disk around a WD.
    The rocky dust disk and their observationally inferred properties are also summarized.
  }
  \label{fig:disk-picture}
\end{figure}

Therefore, we assume the water ice and vapor disk model frame based on the theoretical understandings. Figure~\ref{fig:disk-picture} summarizes our assumed disk properties.
Because $r_{\rm Roche}$ lies within the gravitational radius of gas disks ($\sim 400 R_\odot$), water vapor produced by the sublimation of icy debris is likely to remain gravitationally bound to the central star, forming a disk. Based on the disk evolution simulation of \cite{Okuya+2023}, water vapor undergoes viscous diffusion radially, and upon reaching the snow line (at several 10s of $R_\odot$), recondensation occurs beyond this point. This process can potentially generate an ice particle disk with a width of a few 10s of $R_\odot$ and a mass comprising about a few tens of percent of the initial water vapor \citep{Okuya+2023}. Thus, we assume a water vapor disk inside the snow line ($ \sim 30 R_\odot$) and an ice particle disk beyond the snow line to calculate their thermal emission in Secs.~\ref{subsec:dust-emission} and \ref{subsec:gas-emission}.
\rev{For simplicity, we model the water vapor and ice disks independently without considering their mutual interactions.
However, they are inherently part of the same system, undergoing phase transitions depending on the disk temperature. 
To accurately predict a self-consistent disk structure,} including the mass and spatial distribution of each disk and the grain size of ice, disk evolution simulations incorporating the phase change, drift, and mutual collisions of icy particles would be required. We leave this for future work.

The observationally inferred mass of source pollutant asteroids ranges from $10^{19}$ to $10^{24}$ g \citep{Veras2016}, and we consider the water ice and volatile gas disk mass within this range. Observationally estimated disk masses are also consistent with this range: the rocky dust disk around G29-38 is constrained as at least $5 \times 10^{18}$ g \citep{Ballering+2022}, while the gas disk has not been detected. For SDSS J1228+1040, the mass of the rocky disk and the gas disk of refractory elements is estimated to be $\sim 10^{22}$ g \citep{Brinkworth+2009} and $\sim 10^{21}$ g \citep{Hartmann+2016}, respectively. This suggests that gas disks may be less massive than dust disks.


\subsection{Ice and Dust Disk Emission Model \label{subsec:dust-emission}}
Following Okuya et al. (submitted to ApJ), we calculate \rev{the flux density $F_\nu$ of the emission from rocky dust and water ice at frequency $\nu$} using the radiative equilibrium model for passive disks \citep{Chiang+2001}. We adopt fiducial disk parameters capable of reproducing the spectra observed around G29-38 at wavelengths $< 38\ \mu$m. For the rocky disk, based on previous spectral fitting analyses \citep[][and Okuya et al., submitted to ApJ]{Ballering+2022}, we use a particle size distribution following a power law of $a^{-3.5}$, where $a$ is the dust radius, within $a=1-10\ \mu$m and a total disk mass of approximately $10^{19}$ g, with a surface density profile proportional to $r^{-1}$, where $r$ is the disk radius.
For the water ice disk, we apply a similar particle size distribution but vary the total ice disk mass, $M_{\rm ice}$, from $10^{19}$ to $10^{24}$ g, maintaining a surface density profile proportional to $r^{-1}$. 
For the specific case of the G29-38 icy disk, we set $M_{\rm ice}$ at $\sim 10^{20}$ g to reproduce the slope of the spectrum at wavelengths $> 24\ \mu$m, as observed by Spitzer (see Fig.~\ref{fig:SED_PRIMA_Ice}).

The dust and ice grains are assumed as compact spheres, and their mass absorption opacity is calculated using the publicly available Mie calculation code, OpTool \citep{Carsten+2021}. For rocky dust, we consider dust composed of silicate mixed with metallic Fe and calculate the opacity for this mixed composition (Okuya et al., submitted to ApJ). 
For water ice, we consider pure crystalline H$_2$O, with the optical constants taken from \cite{Warren+2008}, \rev{to estimate the most optimistic detectability scenario. Note that if amorphous ice is present, the 44 $\mu$m ice feature broadens over $40-90\ \mu$m and becomes less distinct \citep[e.g.][]{Smith+1994}, making detection more challenging.}

\subsection{Volatile Gas Disk Emission Model \label{subsec:gas-emission}}
We calculate the integrated line flux, $F$ with the assumed gas compositions, temperatures, and densities (details provided below) using the non-local thermodynamic equilibrium (non-LTE) 1D radiative transfer code, RADEX \citep{VanderTak+2007}.
\rev{We fix the disk inclination at 60$^{\circ}$, approximately the mean inclination for randomly oriented orbits. The disk inclination determines the radial velocity and consequently, spectral line widths. 
While all lines are optically thin in our fiducial parameter case (Sec.~\ref{subsec:results-gas}), if the inclination becomes sufficiently small, the lines can become optically thick, leading to a weakening of their strength.
}
The flux is scaled by multiplying $S / 4\pi d^2$, where $S = \pi(r_{\rm snow}^2 - r_{\rm Roche}^2)$ represents the emitting disk surface area, and $d$ is the distance from Earth to the WD system.

For the gas composition, we assume a single volatile species (e.g., H$_2$O), denoted as $x$, mixed with H$_2$, and vary the mixing ratio $f_{\rm H_2} = n_{\rm H_2}/n_x$, where $n$ is the number density, to explore various compositions without photochemical calculations. Considering that volatile gas disks originate from the icy components of accreted planetary bodies, H$_2$O vapor is expected to be the initial dominant component. However, it may be photo-dissociated by ultraviolet (UV) irradiation from WDs, leading to the formation of other molecules or atoms through further photochemical reactions.
In particular, typical WDs with observed disks have temperatures of approximately 10000 K, and their blackbody radiation peaks around 300 nm, containing a substantial fraction of photons with wavelengths $< 200$ nm capable of dissociating H$_2$O. Therefore, in this study, we include OH and O, in addition to H$_2$O, as candidates for $x$ and consider $f_{\rm H_2} = 10^{-2}, 1, 100$. We leave the incorporation of photochemical calculations for our next work.

We \rev{assume the isothermal gas with a temperature $T_{\rm g}$} and vary $T_{\rm g}$ as a parameter \rev{(see Sec.~\ref{subsec:results-gas} for its influence)}. Specifically, we change it to the equilibrium temperature at $r_{\rm roche}$ (600 K), the equilibrium temperature at $r_{\rm snow}$ (150 K), and an intermediate temperature (300 K). 
 Adopting the isothermal sound speed,
$c_{\rm s} = ({k_{\rm B}T/\mu_{\rm g}})^{1/2}$,
where $\mu_{\rm g}$ is the mean molecular weight of the H$_2$--x mixed gas and $k_{\rm B}$ is the Boltzmann constant, the gas scale height is given by
$H_{\rm g} = c_{\rm s} \Omega^{-1}$, where $\Omega$ is the Keplerian orbital frequency at $r_{\rm gas} = 10 R_{\odot}$, which we adopt as an arbitrary representative value for the gas disk radius.

For simplicity, we assume an gas distributed homogeneously in the radial direction from $r_{\rm Roche}$ to $r_{\rm snow}$, \rev{where the gas surface density $\Sigma_{\rm g}$ is constant with respect to $r$, and in the vertical direction over $H_{\rm g}$.
We can obtain the local gas density of the H$_2$-H$_2$O mixed gas as,
\begin{align}
n_{\rm g} = \frac{\Sigma_{\rm g}}{\sqrt{2\pi}H_{\rm g} \mu_{\rm g}},
\end{align}
where $M_{\rm g} = \int_{r_{\rm Roche}}^{r_{\rm snow}} 2\pi r \Sigma_{\rm g} dr$ and $M_{\rm g}$ is total gaseous mass.}
In our calculation, we vary $M_{\rm g}$ from $10^{19}$ g to $10^{22}$ g (see Sec.~\ref{subsec:disk-overview}).
\rev{This corresponds to $n_{\rm g} \simeq 5\times10^{6}\ {\rm cm^{-3}}$ to $5\times10^{9}\ {\rm cm^{-3}}$ in the fiducial case, assuming $f_{\rm H_2} = 1$ and $T_{\rm g} = 300$ K.}
Note that $M_{\rm g}$ represents the total mass of H$_2$ and $x$. Thus, when $f_{\rm H_2} \lesssim 1$, $x$ contributes most of the mass, whereas when $f_{\rm H_2} \gg 1$, H$_2$ accounts for the majority of the mass.


\subsection{Estimation of Observational Noise \label{subsec:noise}}
To predict the detectability of ice and gas emission spectral features with PRIMA, we estimate the expected observational noise and compare it with the flux of the features.

For the ice disk, since the spectral features of solid materials are relatively broad, low-resolution ($R = \lambda/\Delta \lambda \sim 10$, where $\Delta \lambda$ is the spectral resolution) spectroscopy would be sufficient to resolve these features. Therefore, for ice disk observations, we assume the use of the less observationally expensive low-resolution spectrograph of PRIMA/FIRESS ($R \sim 100$) \rev{\citep{Bradford+2025}}. Additionally, we assume binning of the observed spectra into coarser spectral bins of $R = 10$ to further enhance the signal-to-noise ratios. 
For the low-resolution mode, the expected 5$\sigma$ noise level for a 1-hour observation using PRIMA/FIRESS is $\approx 2 \times 10^{-19}~{\rm W/m^2}$, according to the PRIMA FIRESS Fact Sheet \citep{PRIMA_fact_sheet}. This corresponds to $\approx 600~{\rm \mu Jy}$ at $\lambda=45~\mu$m when scaled to $R=100$. Assuming the noise follows Poisson statistics (i.e., $\sigma_{\rm N} \propto R^{1/2}t^{-1/2}$), we obtain the 1$\sigma$ noise level for the binned spectra as $\sigma_{\rm N} \approx 120(t/1~{\rm hr})^{-1/2}(R/100)^{1/2}~{\rm \mu Jy} \approx 40(t/1~{\rm hr})^{-1/2}(R/10)^{1/2}~{\rm \mu Jy}$.

On the other hand, for the volatile gas disk observations, the molecular lines are so narrow that we assume to use the high-resolution spectrograph of FIRESS ($R \sim 4400$). 
\rev{We do not convolve the calculated volatile gas lines with the instrument's resolution, but this does not significantly affect the estimated line strength since we use the integrated line flux over wavelengths. In addition, for the assumed disk inclination (Sec.~\ref{subsec:gas-emission}), the expected line width is much broader than the spectral resolution of PRIMA, meaning that the peak shape is not affected by the instrumental resolution. }
The expected 5$\sigma$ noise level for a 1-hour observation using the FIRESS high-resolution mode is $\approx 7 \times 10^{-19}~{\rm Wm^{-2}}$ \citep{PRIMA_fact_sheet}. Again, the noise level is assumed to be scaled as $(t/1~{\rm hr})^{-1/2}$. 
\rev{Once the instrument specifications are better defined, detectability should be re-evaluated.}


\section{Results} \label{sec:results}
\subsection{Synthetic Water Ice Emission Spectra and Detectability Prediction}
\begin{figure}
\begin{center}
\includegraphics[bb= 0 0 360 252, height=8.0cm]{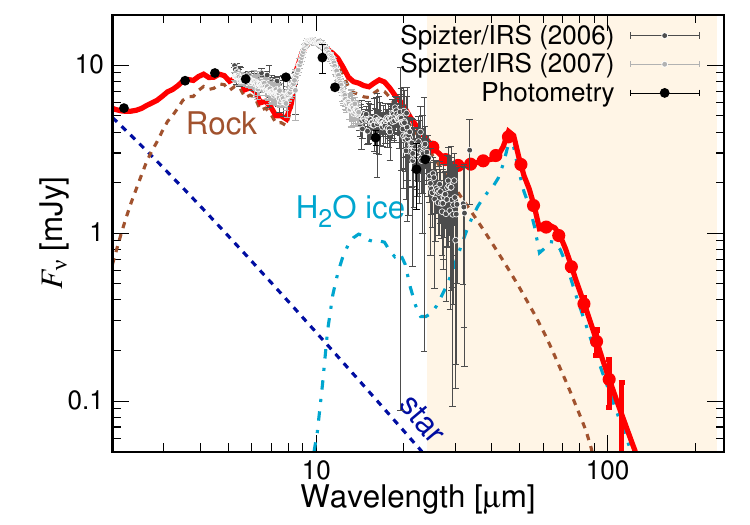}
\end{center}
\caption 
{ \label{fig:SED_PRIMA_Ice}
Synthetic infrared emission spectra of G29-38, incorporating 1$\sigma$ observational noise binned at $R=10$ and assuming 1-hour observations (red line and dots). The contributions of the star, rocky disk, and icy disk are represented by the dark blue dotted line, brown dashed line, and light blue dot-dashed line, respectively.
We plot two Spitzer/Infrared Spectrograph (IRS) spectra sets for comparison: astronomical observation request (AOR) 13828096 (2006), covering $5.2-8.7\ \mu$m and $14.0-38\ \mu$m, and AOR 22957568 (2007), covering $5.2-14.5\ \mu$m. The IRS spectra are obtained from the Combined Atlas of Sources with Spitzer IRS Spectra \citep{Lebouteiller+2011} online database. We also include photometry data compiled in \citep{Ballering+2022}.
The orange shaded region marks wavelength ranges covered by PRIMA/FIRESS ($24-235\ \mu$m). } 
\end{figure}

First, we demonstrate the calculated infrared spectra from rocky and icy disks, using disks around G29-38 as an example, in Fig.~\ref{fig:SED_PRIMA_Ice}. 
For the spectra generation, we use the stellar parameters of G29-38 \citep{Subasavage+2017}.
The emission from the hot rocky disk dominates the spectrum at $\lambda \lesssim 20\ {\rm \mu m}$, whereas emission from the icy disk dominates at $\lambda \gtrsim 20\ {\rm \mu m}$. This is because the temperature of the icy disk is as low as $\sim 150$ K and lower-temperature ice emits more brightly at longer wavelengths. Notably, water ice features emerge at 44 $\mu$m and 62 $\mu$m. In particular, the prominent peak at 44 $\mu$m could be especially useful for detecting ice disks with PRIMA observations.

\rev{
We note that our synthetic spectrum does not perfectly match the observed one. 
The model by \citep{Reach+2009}, which incorporated a more detailed mineralogical composition with water ice,
better explains the spectral slope at $\lambda \gtrsim 24 \,\mu$m. 
Yet, the Spitzer data at long wavelengths suffer from a low signal-to-noise ratio. A reanalysis of high-precision spectra from PRIMA, incorporating a more comprehensive set of candidate materials into our disk model, will provide stronger constraints on the amount of water ice in the G29-38 disk.}

Next, we predict the detectability of ice disks using the promising 44 $\mu$m ice feature. \rev{To quantify the significance of the feature relative to observational noise, we introduce the Feature-to-Noise ratio (F/N) as a metric, defined as (see \citep{Okuya+2020} for a more detailed explanation),
\begin{equation}
{\rm F/N} = \frac{F_{\nu} (\lambda_{\rm peak})- F_{\nu} (\lambda_{\rm floor})}{\sqrt{{\sigma_{\rm N} (\lambda_{\rm peak})}^{2}+{\sigma_{\rm N} (\lambda_{\rm floor})}^{2}}},
\label{eq:FN}
\end{equation}
where $\lambda_{\rm peak}$ is the wavelength at which the ice feature peaks, and $\lambda_{\rm floor}$ is the wavelength at which the flux density reaches a local minimum just outside the feature.
Here, the numerator represents the relative flux density of the feature, while the denominator accounts for the observational noise at $\lambda = \lambda_{\rm peak}$ and $\lambda_{\rm floor}$. 
In brief, a higher F/N value indicates a more detectable feature.
}
For systems at various distances from Earth, we vary $M_{\rm ice}$ while keeping other disk parameters fixed to produce theoretical emission spectra and calculate the corresponding F/N.

\begin{figure}
\begin{center}
\begin{tabular}{c}
\includegraphics[bb=0 0 528 341, height=7.5cm]{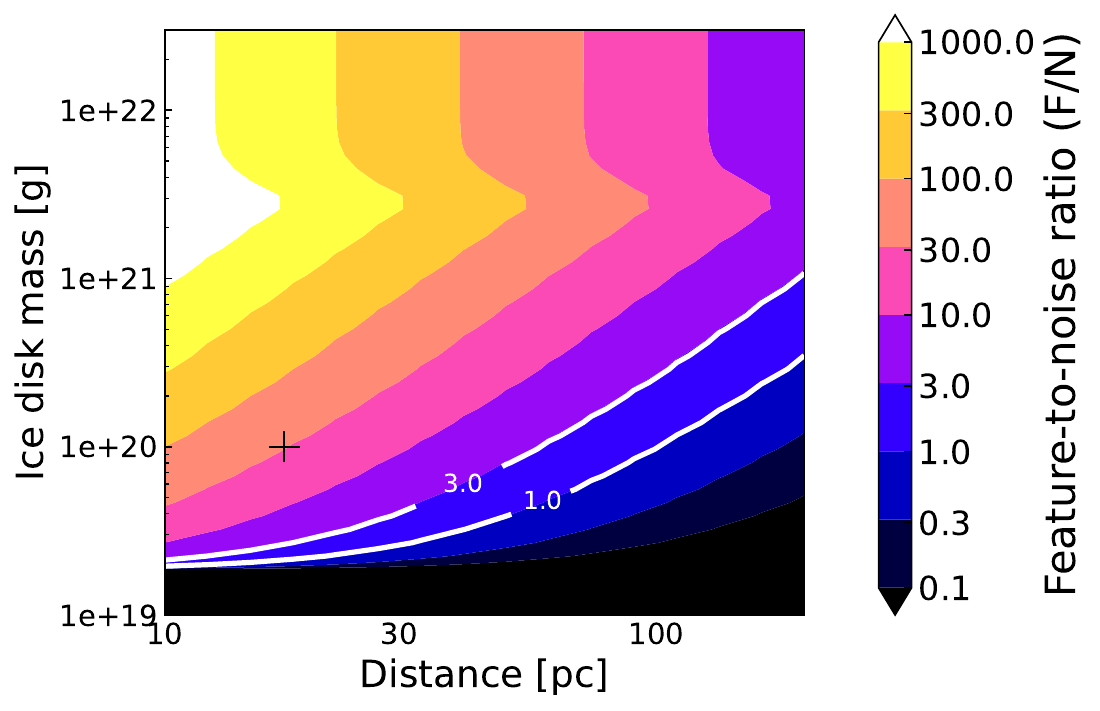}
\end{tabular}
\end{center}
\caption 
{ \label{fig:FNmap_w_rock}
Feature-to-noise ratios (F/N) for the 44 $\mu$m water ice feature from a 1-hour observation with PRIMA/FIRESS as a function of the distance to the system, $d$, and the water ice dust mass $M_{\rm ice}$. The white contours mark ${\rm F/N} = 1$ and 3. The balck cross symbol indicates the position of G29-38.
} 
\end{figure}

Figure~\ref{fig:FNmap_w_rock} presents the calculated F/N ratio as a function of the distance to the WD system and $M_{\rm ice}$. For G29-38, if $M_{\rm ice} = 10^{20}$ g, which we adopt for reproducing the observed spectra in Fig.~\ref{fig:SED_PRIMA_Ice}, PRIMA/FIRESS would be able to detect the water ice feature with F/N $\sim$ 30. Furthermore, for an ice disk with the same mass, the disk around WDs at $\lesssim 60$ pc could be \rev{marginally} detectable with $\rm F/N > 3$. 
For $M_{\rm ice} \lesssim 2 \times 10^{19}$ g, the emission from the rocky disk surpasses that from the icy disk, dimming the ice feature in the disk spectra. On the other hand, for $M_{\rm ice} \gtrsim 3 \times 10^{21}$ g, the disk becomes optically thick, causing the feature intensity to start saturating. 
These boundary masses depend on the assumed particle size.
We note that the apparent discontinuity in the F/N contours at $M_{\rm ice} \gtrsim 3 \times 10^{21}$ g is artificial, arising from the approximate treatment of the vertical optical thicknesses in the two-layer disk model (Okuya et al., submitted to ApJ). 

\subsection{Simulated Volatile Gas Emission Lines and their Parameter Dependences} \label{subsec:results-gas}

\begin{figure}
\centering
    \includegraphics[bb= 0 0 720 540, height=12.0cm]{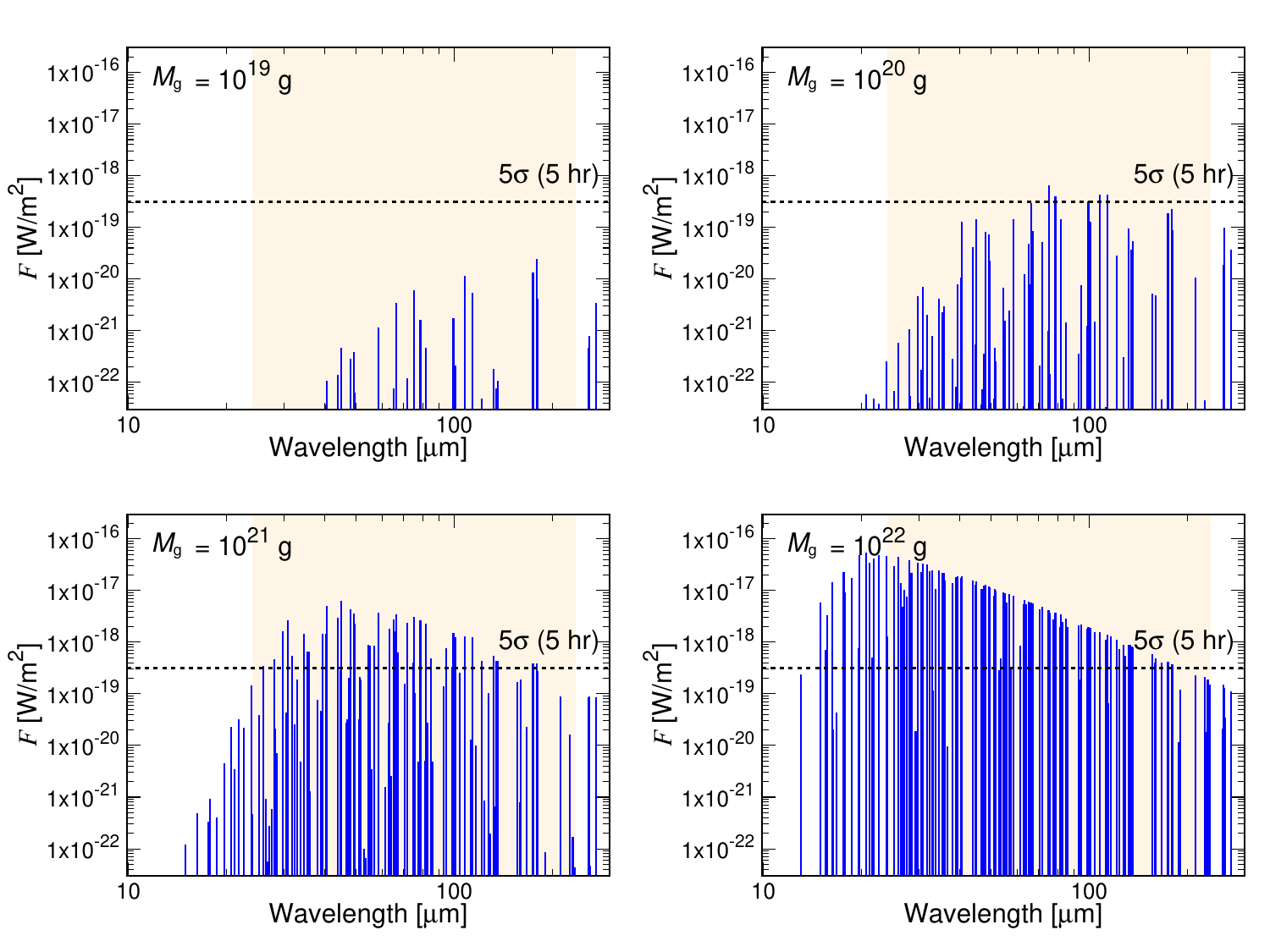}
\caption{
Synthetic emission lines of water vapor with the fiducial disk parameters for various gas masses, $M_{\rm g}$. For comparison, the expected 5-$\sigma$ observational noise for a 5-hour observation with PRIMA/FIRESS is shown with the black dotted line. The orange shaded region marks the wavelength range covered by PRIMA/FIRESS ($24-235\ \mu$m).
}
\label{fig:gas-mass}
\end{figure}

First, we present the synthetic emission lines of water vapor with the fiducial disk parameters ($f_{\rm H_2}=1$, $T_{\rm g}=300$ K) in Fig.~\ref{fig:gas-mass}. We use the distance to G29-38, $d=17.5$ pc. Numerous rotational lines appear at wavelengths $\gtrsim 30\ \mu$m, which fall within the PRIMA/FIRESS observational band. The line flux increases with gas mass until the lines become optically thick ($M_{\rm g} \gtrsim 10^{22}$ g). We assess detectability by simply comparing the line flux with the expected 5-$\sigma$ observational noise for the high-resolution mode of PRIMA/FIRESS. For a system at $\sim 20$ pc, a 5-hour PRIMA observation would be able to detect volatile gas with $M_{\rm g} \gtrsim 10^{20}$ g, potentially enabling the detection of most estimated disk masses.

\begin{figure}
\centering
    \includegraphics[bb= 0 0 720 540, height=12.0cm]{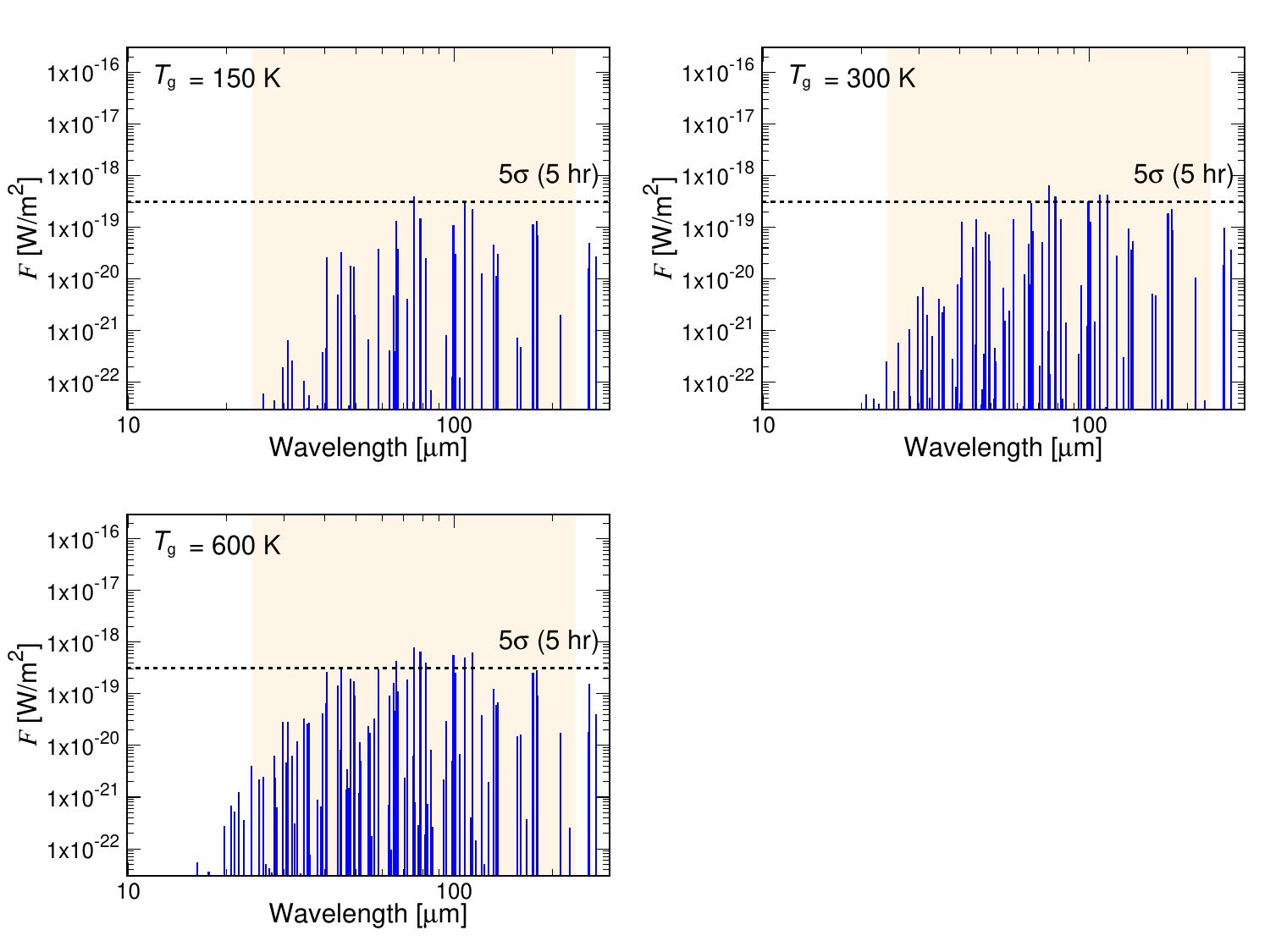}
\caption{Same as Fig.~\ref{fig:gas-mass}, except for varying the gas temperature, $T_{\rm g}$. The total gas mass is fixed at $10^{20}$ g.}
\label{fig:gas-temperature}
\end{figure}

Next, we examine the impact of $T_{\rm g}$ on line observability. In general, higher temperatures tend to produce stronger emission lines and increase the number of possibly detectable lines. 
The line flux depends directly on $T_{\rm g}$ for the higher-mass disk case ($M_{\rm g} \gtrsim 10^{21}$ g), where lines are in LTE ($T_{\rm ex} = T_{\rm g}$). We find that as long as $T_{\rm g} \ge 150$ K, some optically thick lines remain detectable above the 5-$\sigma$ observational noise level in this case.
In contrast, the effect of temperature is reduced in the lower-mass disk case ($M_{\rm g} \lesssim 10^{20}$ g).
Figure \ref{fig:gas-temperature} shows the emission spectra for gas with fixed $M_{\rm g}=10^{20}$ g but varying $T_{\rm g}$.
In this case, lines are in non-LTE, meaning the excitation temperatures decoupled from the kinetic gas temperature, $T_{\rm g}$.
Thus, temperature dependence may not be a critical factor for detectability of some lines within our assumed parameter space.

\begin{figure}
\centering
    \includegraphics[bb= 0 0 720 540, height=12.0cm]{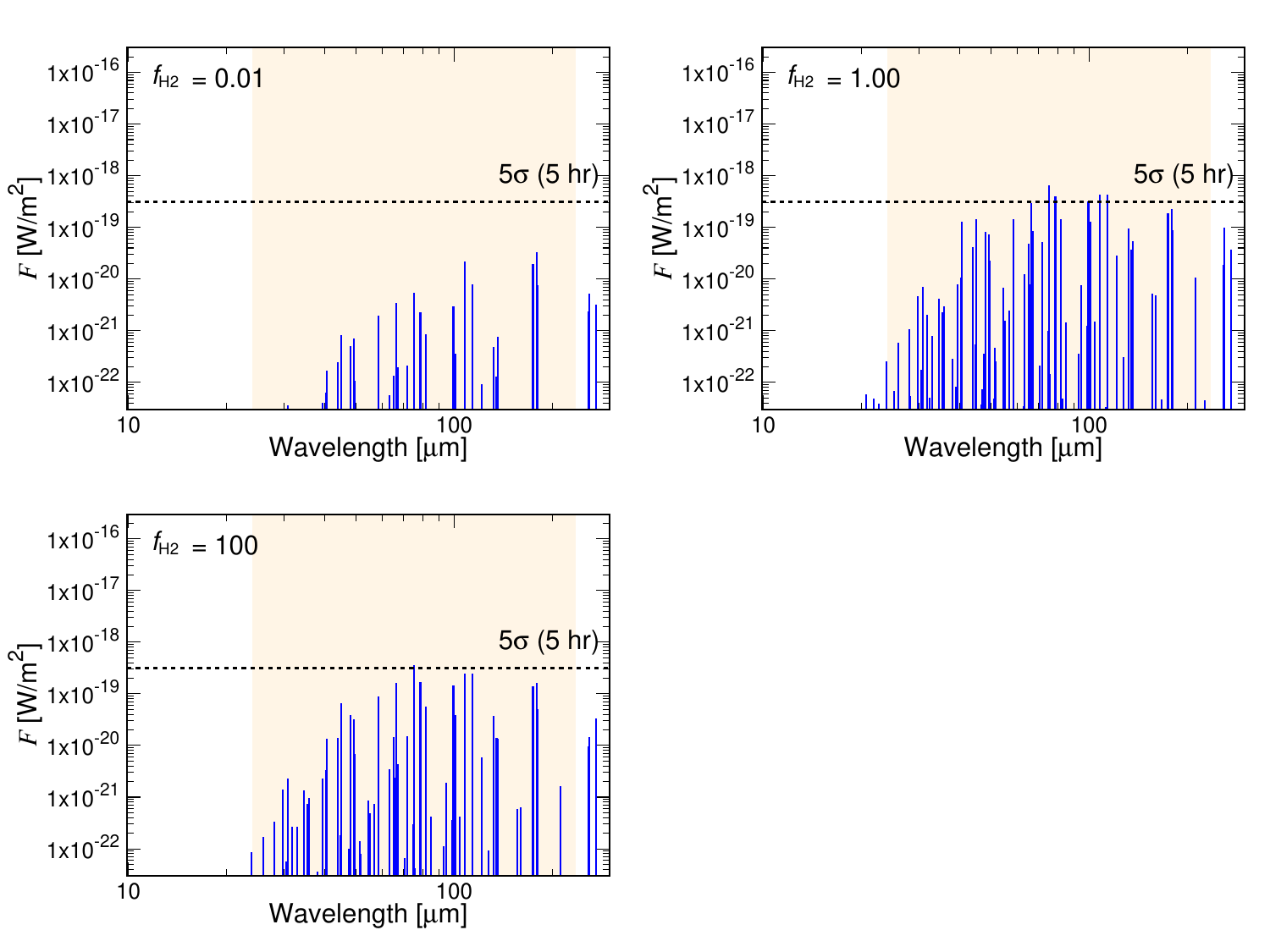}
\caption{
Same as Fig.~\ref{fig:gas-mass}, except for varying the number density mixing ratio of H$_2$ to H$_2$O, $f_{\rm H_2}$. The total gas mass is fixed at $10^{20}$ g.
}
\label{fig:H2mixing}
\end{figure}

We also investigate the dependence of line observability on $f_{\rm H2}$. Figure \ref{fig:H2mixing} shows the calculated water vapor emission lines for different $f_{\rm H2}$ values with the total gas mass fixed at $M_{\rm g}=10^{20}$ g. 
For $f_{\rm H2} < 1$, the line flux decreases as $f_{\rm H2}$ decreases because the lower H$_2$ density causes the H$_2$O lines to be in non-LTE, making detection more challenging. 
%
For $f_{\rm H2} \ge 1$, the difference in line flux between $f_{\rm H2} = 1$ and $f_{\rm H2} = 100$ is slight, even though the H$_2$O column density is an order of magnitude lower for $f_{\rm H2} = 100$ (due to the fixed total gas mass). 
This is because the higher H$_2$ density compensates for the lower H$_2$O density through collisional excitation, resulting in line emission as strong as in the case of $f_{\rm H2} = 1$. 
\rev{However, this compensation does not lead to a complete degeneracy between $f_{\rm H_2}$ and the H$_2$O mass. The former affects the LTE/non-LTE conditions of the lines, while the latter influences their optical thickness. 
Therefore, if multiple lines, including both optically thin and (marginally) thick lines in both LTE and non-LTE conditions, are detected, both quantities can, in principle, be constrained independently.}
Thus, in addition to disk mass, the density of the background gas significantly affects observability and is crucial for interpreting observations. 
Chemical reaction calculations would help predict $f_{\rm H_2}$ from a theoretical perspective, which we leave for our next work.



\begin{figure}
\centering
    \includegraphics[bb= 0 0 720 252, height=5.5cm]{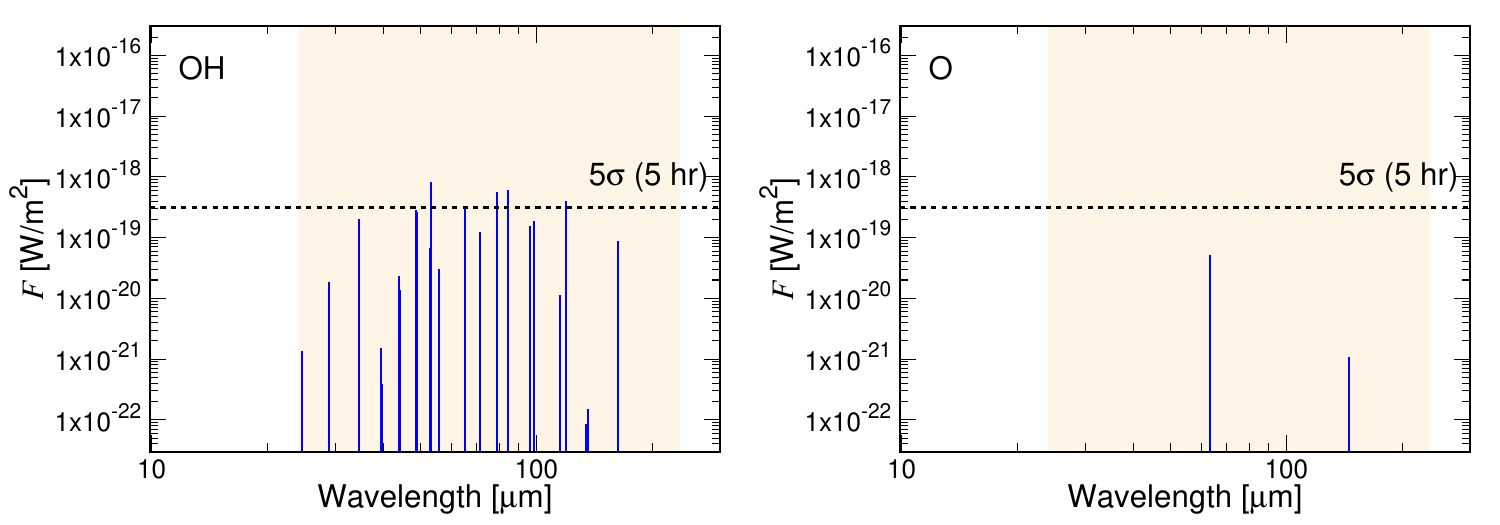}
\caption{Same as Fig.~\ref{fig:gas-mass}, except for OH and O gas. The total gas mass is fixed at $10^{20}$ g.
}
\label{fig:OHs}
\end{figure}

The emission spectra of OH and O with fiducial disk parameters and $M_{\rm g} = 10^{20}$ g are shown in Fig.~\ref{fig:OHs}. The OH lines appear with the intensity similar to those of H$_2$O within the PRIMA/FIRESS wavelength range, suggesting that OH may be detectable. In contrast, O shows fewer lines, and their intensity is significantly weaker compared to both H$_2$O and OH, indicating that detection would be challenging unless a substantial mass is present.

\section{Discussion}\label{sec:discussion}
\subsection{Dependence on Particle Size}

So far, we have assumed an ice grain size of $1-10\ \mu$m. Although the ice grain size remains observationally unconstrained, we examine the effects of grain size on icy disk detectability, particularly focusing on the F/N ratio.
To this end, for a disk with a mass of $10^{20}$ g located at 10 pc, we consider ice grains with a single size and vary their radius from 0.1 $\mu$m to 30 $\mu$m while keeping other disk parameters fixed, then calculate the corresponding F/N ratio. The results are as follows: F/N $\sim 130$ for $a=0.1\ \mu$m, F/N $= 100$ for $a=1\ \mu$m, F/N $= 10$ for $a=10\ \mu$m, and F/N $= 0$ for $a=30\ \mu$m. 
Note that with a particle size distribution, for example, following $\propto a^{-3.5}$ with $a=1-30 \mu$m, the ice feature remains detectable with F/N$\sim 30$.

The F/N ratio remains nearly constant for $a \ll 2\pi/\lambda$, where we substitute the wavelength of the water ice feature, $\lambda = 44\ \mu$m. In contrast, 
the F/N ratio sharply decreases as the particle radius increases, particularly when $a \gtrsim 2\pi/\lambda$.
This is because spectral features become less sharp for larger grains and eventually disappear.
Therefore, as the grain size increases, the region where F/N $> 3$ is expected to shift to disks with larger ice mass and closer distance.

\subsection{Target Candidates for Water Vapor and Ice Disk Detections}
We here discuss the number of potential targets suitable for observing water vapor and ice disks, respectively based on our assessments in Sec.~\ref{sec:results}.  
Water vapor with a disk mass of $10^{20}$ g, which is roughly the estimated lower limit of circumstellar disk mass, could potentially be detected with a 5-hour PRIMA/FIRESS observation if a WD is located within 20 pc.
Within this distance, 75 single WDs are known \citep{MWDD}. Among them, 19 currently show signs of metal pollution, and 2 of these exhibit infrared excess. Therefore, these 19 WDs would be optimal targets for observing water vapor.
In addition to the PRIMA/FIRESS wavelength range, vibrational and rotational lines of water vapor are abundantly present across near-infrared to mm wavelengths.
Observations using the James Webb Space Telescope (JWST) and Atacama Large Millimeter/submillimeter Array (ALMA) may also be effective in detecting these emission lines.

On the other hand, the wavelength range around 44 $\mu$m, where the water ice feature is located, can only be observed by PRIMA.
A water ice disk of the same mass within 60 pc could be detectable with a 1-hour PRIMA/FIRESS observation. Within 60 pc, potential targets are ten times more numerous than those for gas detection, with $\sim 210$ metal-polluted WDs, \rev{16} of which show detectable infrared disk emission \citep{MWDD}.
\rev{Among these, one was recently discovered through the JWST Cycle 2 disk survey \citep{Farihi+2025}. This survey, using  Mid-Infrared Instrument (MIRI) Low Resolution Spectroscopy (LRS), achieved a disk detection rate approximately three times higher than that of Spitzer, as reported by \citep{Farihi+2025}. 
While most metal-polluted white dwarfs are expected to host disks, the detection rate so far has remained around 10\% \citep{Rocchetto+2015}. Therefore, future JWST MIRI observations will likely reveal new dusty white dwarfs that have not been previously detected but could be promising targets for PRIMA.
}




\section{Conclusion} \label{sec:conclusion}
We have investigated the detectability of water ice and vapor disks around WDs, which are thought to originate from the icy components of accreted planets and/or asteroids using future far-infrared PRIMA observations. We employed a simple dust, ice, and gas disk emission model and assumed disk parameter ranges inferred from both theoretical and observational studies to produce synthetic observational spectra.

We have found that the emission from cold ice disks would dominate the emission from hot rocky dust disks in the wavelength range covered by PRIMA/FIRESS, and in particular, 44 $\mu$m water ice feature would be promising (Fig.~\ref{fig:SED_PRIMA_Ice}). Based on the expected observational noise of the low-resolution spectrographs of PRIMA/FIRESS, we have quantified the detectability of the 44 $\mu$m water ice feature in disk emission spectra binned at $R=10$. For WDs within 60 pc, 1-hour PRIMA observations would be capable of detecting water ice with a total mass larger than $10^{20}$ g (Fig.~\ref{fig:FNmap_w_rock}), which may represent the potential lower limit of circumstellar disk mass.

We have also found that the rotational emission lines from water vapor are abundant within the PRIMA/FIRESS band. The line flux is most sensitive to the total mass of the gas disk (Fig.~\ref{fig:gas-mass}) and less sensitive to the gaseous temperature (Fig.~\ref{fig:gas-temperature}). For WDs located within 20 pc, 5-hour observations could potentially detect water vapor lines in disks with a total gas mass larger than $10^{20}$ g, covering most of the estimated disk masses.
However, the mixing ratio of H$_2$ to H$_2$O significantly affects the line flux when the mixing ratio smaller than unity (Fig.~\ref{fig:H2mixing}), as it changes the fraction of non-LTE and LTE lines. Incorporating photochemical simulations to determine the composition of volatile gas and preparing for future spectral analyses will be left for our next work.

Finally, we have discussed candidate targets that meet the observation conditions: 19 metal-polluted WDs within 20 pc and $\sim210$ within 60 pc are identified as optimal targets for gas and dust observations with PRIMA, respectively. In particular, G29-38, located at 17.5 pc, has shown potential signs of a water ice feature in its observed infrared spectrum, making it a primary target.

Measuring the abundance of volatile components in circumstellar WD disks would enable constraints on the volatile content of accreted planets/asteroids, which has been difficult to infer from WD atmospheric observations, unlike refractory elements. This would represent the first opportunity to reveal the full elemental composition of exoplanetary material. The abundance of volatile elements provides crucial insights into the temperature conditions during planet formation, offering constraints on the formation time and location of planets in protoplanetary disks \citep[e.g.,][]{Marboeuf+2014,Eistrup+2018}.



\subsection*{Disclosures}
We declare no relevant financial interests in the manuscript and no other potential conflicts of interest to disclose.

\subsection* {Code, Data, and Materials Availability} 
This study utilizes publicly available codes, OpTool \citep{Carsten+2021}, RADEX \citep{VanderTak+2007}, matplotlib \citep{Hunter2007}, and gnuplot \citep{gnuplot}. The code for calculating the SED of dust disks will be made available on GitHub upon the publication of the corresponding paper (Okuya et al., submitted to ApJ).


\subsection* {Acknowledgments}
The authors thank the PRIMA-J science team. We also thank the referees for their constructive comments and suggestions that helped improve this paper. This work was supported by Japan Society for the Promotion of Science (JSPS) KAKENHI (Grant No. JP22KJ3094).



\vspace{2ex}\noindent\textbf{Ayaka Okuya} is a JSPS Postdoctoral Research Fellow at the National Astronomical Observatory of Japan. She received her PhD in earth and planetary sciences from the Institute of Science Tokyo in 2022. \rev{Her research focuses on exoplanet properties, including bulk composition and surface environments. She has been working on modeling debris disks around white dwarfs as a key to measuring planetary composition.} She is a member of the PRIMA-J star and planet formation science team.

\vspace{2ex}\noindent\textbf{Hideko Nomura} is a professor at Division of Science, the National Astronomical Observatory of Japan. She received her PhD in astronomy from Kyoto University in 2001. She has been working on modelling physical and chemical structure of protoplanetary disks, and also involved in observational projects, such as ALMA (Atacama Large Millimeter/submillimeter Array), recently. She is a co-lead of the PRIMA-J star and planet formation science team.



\listoffigures
\listoftables

\end{spacing}
\end{document}